\documentclass[12pt,epsfig,a4]{article} 
\newcommand{\vs}{\vspace{-0.25cm}}
\usepackage{amsmath,graphicx}
\begin{document} 

\begin{center}
{\Large{\bf Third-order particle-hole ring diagrams with
contact-interactions and one-pion exchange}\footnote{This work 
has been supported in part by DFG and NSFC (CRC110).}  }  

\bigskip

N. Kaiser \\
\medskip
{\small Physik-Department T39, Technische Universit\"{a}t M\"{u}nchen,
   D-85748 Garching, Germany}
\end{center}
\medskip
\begin{abstract} The third-order particle-hole ring diagrams are 
evaluated for a NN-contact interaction of the Skyrme type. The 
pertinent four-loop coefficients in the energy per particle $\bar E(k_f)
\sim k_f^{5+2n}$ are reduced to double-integrals over cubic expressions in 
euclidean polarization functions. Dimensional regularization of divergent 
integrals is performed by subtracting power-divergences and the validity 
of this method is checked against the known  analytical results at 
second-order. The complete ${\cal O}(p^2)$ NN-contact interaction is obtained
by adding two tensor terms and their third-order ring 
contributions are also calculated in detail. The third-order ring energy 
arising from long-range $1\pi$-exchange is computed and it is found that 
direct and exchange contributions are all attractive. The very large 
size of the pion-ring energy, $\bar E(k_{f0})\simeq -92\,$MeV at saturation 
density, is however in no way representative for that of realistic chiral 
NN-potentials. Moreover, the third-order (particle-particle and 
hole-hole) ladder  diagrams are evaluated with the full ${\cal O}(p^2)$ 
contact interaction and the simplest three-ring contributions to the 
isospin-asymmetry energy $A(k_f)\sim k_f^5$ are studied. 
\end{abstract}

\section{Introduction and summary}
Due to the progress in constructing nuclear forces within chiral effective field 
theory \cite{epel,evgeny,machleidt} and the advances in many-body techniques, 
new paths have opened to perform systematically improvable calculations of nuclear 
many-body systems. Infinite (isospin-symmetric) nuclear matter and pure neutron 
matter have been studied extensively based on chiral two- and three-nucleon 
low-momentum interactions within various many-body frameworks, such as many-body 
perturbation theory \cite{achprl,achim,drisch,wellen,phhole}, an in-medium chiral 
perturbation approach \cite{review}, the selfconsistent Green's function method 
\cite{carb1,carb2}, the Brueckner-Hartree-Fock approach \cite{isaule}, and quantum 
Monte-Carlo simulations \cite{lynn}. When including the (leading) chiral 
3N-force in the form of a density-dependent effective NN-interaction, calculations 
at least up to second order are necessary in order to achieve reasonable saturation 
properties and agreement with empirical bulk quantities of nuclear matter. Several 
works \cite{achim,drisch,carb2} have also studied the importance of the third-order
particle-particle and hole-hole ladder contributions, but the third-order 
particle-hole ring diagram is often neglected due to its more complicated momentum 
and spin recouplings when implementing the NN-potential in terms of partial-wave 
matrix-elements, which significantly increases the computational costs. In 
ref.\cite{phhole} it was found that the third-order particle-hole ring diagram 
gives a contribution of about $(1-2)\,$MeV in isospin-symmetric nuclear matter
at $\rho_0 \simeq 0.16\,$fm$^{-3}$ when computed from low-resolution chiral 
NN-potentials. This size of the three-ring energy per particle has been confirmed 
recently in ref.\cite{jeremy}, where the extensive numerical computations involving 
multiple partial-wave sums and momentum-space integrations have been benchmarked 
against semi-analytical evaluations for test-interactions of the one-boson 
exchange type.

The purpose of the present paper is to calculate the third-order particle-hole ring 
diagrams for relatively simple two-body interactions that just make feasible an 
analytical treatment. A suitable candidate for such an interesting and worthwhile 
exploration within many-body perturbation theory is the NN-contact interaction of 
the Skyrme type. The latter is widely and successfully used for non-relativistic 
nuclear structure calculations of medium-mass and heavy nuclei \cite{bender,stone}. 
Actually, the purely phenomenological Skyrme force should be viewed as to provide a 
convenient parametrization of the nuclear energy density functional on which the 
self-consistent mean-field treatment can be based. Following the current efforts 
to build new functionals from many-body techniques, the second-order contributions 
arising from the Skyrme NN-contact interaction in nuclear matter have been 
calculated in ref.\cite{sk2}. The second-order results obtained for various nuclear 
matter quantities consist of even powers of the Fermi momentum $k_f$ multiplied by 
products of the Skyrme parameters and a numerical coefficient of the form 
$\ln 2+r_j$, where $r_j$ is some rational number. This line of attempt has been 
continued recently by Moghrabi in ref.\cite{kassem}, where the 
next-to-next-to-leading order Skyrme interaction (quartic in momenta) has been 
considered and nuclear bulk quantities have been calculated from it up to 
second order. In view of these developments it is one particular aim of the 
present work to derive the (complete) third-order contributions from ring and 
ladder diagrams for the next-to-leading order Skyrme interaction (quadratic in 
momenta). The general NN-contact interaction of order ${\cal O}(p^2)$ is readily 
obtained by adding two tensor terms and their third-order contributions 
are evaluated together with all possible interference terms.

Let us remind that the third-order contributions known so far in the literature 
constitute two parts in the low-density expansion for an interacting many-fermion 
system \cite{lowdexp}, which read: 
\begin{equation} \bar E(k_f)^{\rm 3-ring} = (1-g)(g-3){a^3 k_f^5 \over M\pi^4} 
\cdot 2.7950523\,, \qquad
\bar E(k_f)^{\rm 3-lad} = (1-g){a^3 k_f^5 \over M\pi^3} \cdot 1.1716223\,.
\end{equation} 
These results for the energy per particle $\bar E(k_f)$ derive from a 
(momentum-independent) contact-interaction proportional to the s-wave scattering 
length $a$ (where $a>0$ corresponds to attraction) and $g$ is the spin-degeneracy 
factor entering the relation between the density $\rho = g\,k_f^3/6\pi^2$ and 
the Fermi momentum $k_f$. The large fermion mass is denoted by 
$M$. In the course of the present work it will also become clear, how the numerical 
coefficients can be computed with the given accuracy.

The present paper is organized as follows. In section 2 we prepare the evaluation 
of third-order particle-hole ring diagrams by introducing the antisymmetrized 
Skyrme NN-contact interaction which allows to treat in one single step 
direct and exchange-type contributions together. For the momentum-dependent 
interaction product resulting from the spin- and isospin-traces, three of the four 
loop-integrations can be factorized and solved in terms of cubic expressions in 
euclidean polarization functions. The results for the three-ring energy per 
particle of isospin-symmetric nuclear matter and pure neutron matter are given in 
section 3, where the focus lies on the computation of the pertinent four-loop 
coefficients with high numerical accuracy. Dimensional regularization of divergent 
integrals is performed by subtracting power-divergencs from the integrand. The 
validity of this practical method is checked against the known analytical results at 
second order. In section 4 the general ${\cal O}(p^2)$ NN-contact interactions is 
completed by introducing two additional tensor interactions and the corresponding 
third-order ring contributions are evaluated together with all possible 
interference terms. Section 5 is devoted to a semi-analytical evaluation of the
third-order particle-hole ring diagrams with the long-range one-pion exchange 
interaction.  It is found that exchange-type corrections are smaller 
than the  direct three-ring contribution, but these add up coherently to a very 
large attraction of $\bar E(k_{f0}) \simeq -92\,$MeV. Finally, in section 6 the 
contributions from the third-order (particle-particle and hole-hole) ladder diagrams
are evaluated for the general ${\cal O}(p^2)$ NN-contact interaction. Moreover, in 
the appendix the three-ring contributions to the isospin-asymmetry energy $A(k_f) 
\sim k_f^5$ as they arise from a contact-interaction involving two scattering 
lengths $a_s, a_t$ are studied.

\section{Ring diagrams and antisymmetrized Skyrme-interaction} 
\begin{figure}[h!]
\begin{center}
\includegraphics[scale=0.6,clip]{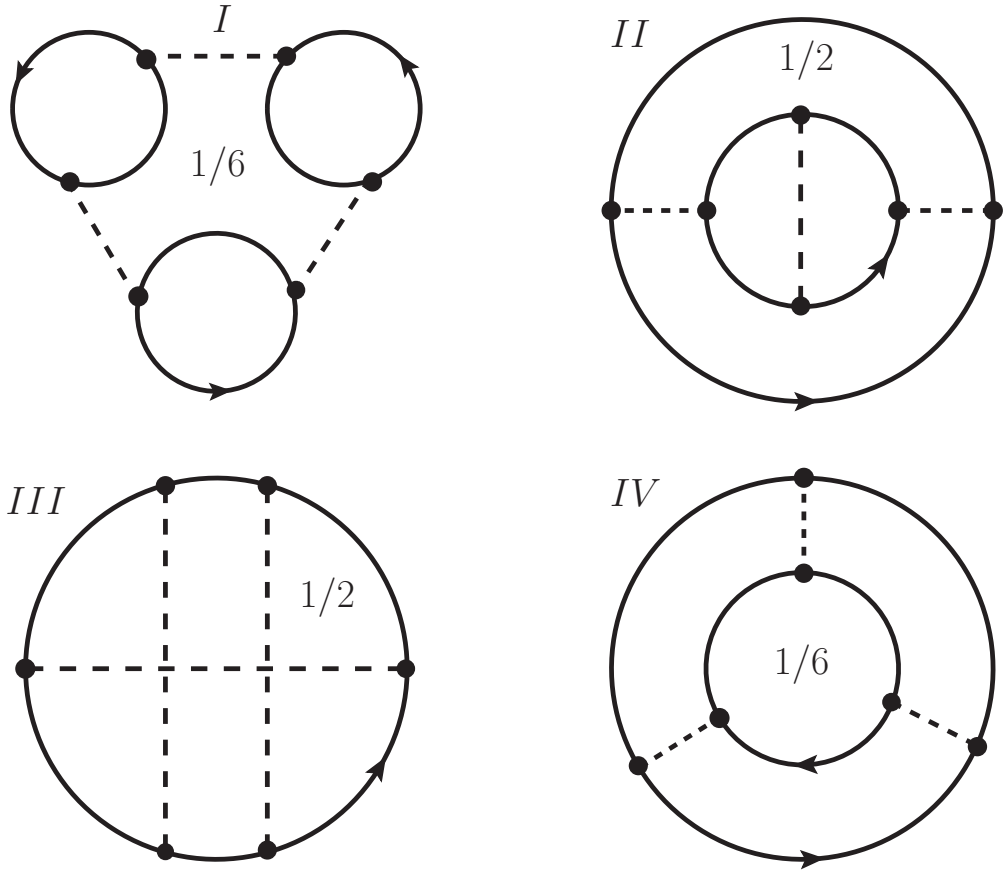}
\end{center}
\vspace{-.6cm}
\caption{Third-order particle-hole ring diagrams: $I = dir^3,\, II = 
-dir^2\,exc,\, III = dir\,exc^2$ and $IV =  -exc^3$. The corresponding symmetry 
factors $1/6,\, 1/2,\, 1/2$ and $1/6$ are indicated. The dashed line represents 
the Skyrme contact-interaction $V_{\rm Sk}$ or one-pion exchange.}
\end{figure}
The four topologically distinct ring diagrams representing the third-order 
particle-hole contribution to the energy density are shown in Fig.\,1, organized 
according to the number of direct and exchange interactions. Together with 
their symmetry factors $1/6,1/2,1/2,1/6$ and signs from the number of closed 
fermion-lines these diagrams result formally from the expansion of 
$-1/6(dir-exc)^3$. Therefore, the three exchange-type diagrams are automatically 
included by employing the antisymmetrized two-body interaction $dir-exc$. In the 
case of the Skyrme NN-contact interaction \cite{sk2}, the antisymmetrization 
operator leaves unchanged the momentum-dependence:
\begin{eqnarray} && V_{\rm Sk} -P_\sigma P_\tau V_{\rm Sk}\big|_{\vec q_{\rm out}\to -\vec 
q_{\rm out}} = (1-P_\sigma P_\tau)\Big\{t_0(1+x_0 P_\sigma) +{t_1 \over 2}(1+x_1 P_\sigma) 
(\vec q_{\rm out}^{\,2}+ \vec q_{\rm in}^{\,2})\Big\}\nonumber \\ &&  \qquad\qquad\quad
+(1+P_\sigma P_\tau)\,t_2(1+x_2 P_\sigma)\, \vec q_{\rm out}
\!\cdot\! \vec q_{\rm in}+ (1+P_\tau) iW_0(\vec \sigma_1+\vec \sigma_2)\!\cdot\!
(\vec q_{\rm out}\!\times\!\vec q_{\rm in})\,.\end{eqnarray}
Here, $P_\sigma=(1+\vec \sigma_1\cdot\vec \sigma_2)/2$ and $P_\tau=(1+
\vec \tau_1\cdot\vec \tau_2)/2$ are the spin- and isospin-exchange operators,
while  $\vec q_{\rm in} = (\vec p_1-\vec p_2)/2$ and  $\vec q_{\rm out} = ({\vec p_1}\,
\!\!'-\vec p_2\,\!\!')/2$ denote half of the momentum difference in the initial 
and final state, respectively. Note that the anti-symmetrized spin-orbit 
contact-interaction $\sim W_0$ has been simplified by using the relation 
$\vec \sigma_1\cdot\vec \sigma_2(\vec \sigma_1+\vec \sigma_2) =\vec \sigma_1 
+\vec \sigma_2$. The three closed nucleon-lines of the third-order ring diagram 
$I$ in Fig.\,1 are associated with triple spin- and isospin traces, 
which are readily computed with the help of the following master formula: 
\begin{eqnarray} && {1 \over 64} {\rm tr}_1{\rm tr}_2{\rm tr}_3\big\{
(A+ B \vec\sigma_1\!\cdot\!\vec \sigma_2+ C \vec\tau_1\!\cdot\!\vec \tau_2+ 
D\vec  \sigma_1\!\cdot\!\vec \sigma_2\vec \tau_1\!\cdot\!\vec \tau_2)(A'+ B'
\vec \sigma_2\!\cdot\!\vec \sigma_3+ C'\vec  \tau_2\!\cdot\!\vec \tau_3+ D'
\vec \sigma_2\!\cdot\!\vec \sigma_3 \vec\tau_2\!\cdot\!\vec \tau_3)\nonumber 
\\ && \times (A''+ B''\vec  \sigma_3\!\cdot\!\vec \sigma_1+ C'' \vec \tau_3\!
\cdot\!\vec \tau_1+D'' \vec \sigma_3\!\cdot\!\vec \sigma_1 \vec\tau_3\!\cdot
\!\vec \tau_1)\big\}=AA'A''+3BB'B''+3CC'C''+9DD'D''\,.\end{eqnarray}
Applying the decomposition into $A,B,C,D$ to three antisymmetrized Skyrme 
NN-contact interactions and treating separately the spin-orbit term, one obtains 
the following expression for the interaction product:
\begin{eqnarray} &&12t_0^3(1-6x_0^2) +9t_0^2 t_1(1-2x_0^2-4x_0x_1) 
\big(\vec l^{\,\,2}_{12}+\vec q^{\,2}\big)  +9t_0^2 t_2\big[5+4x_2+2x_0^2(1+2x_2)
\big] \big(\vec l^{\,\,2}_{12}-\vec q^{\,2}\big) \nonumber\\ && +{9\over 4}t_0 
t_1^2(1-4x_0 x_1-2x_1^2) \big(\vec l_{12}^{\,\,2}\,\vec l_{13}^{\,\,2}+2\vec 
l_{12}^{\,\,2}\,\vec q^{\,2}+\vec q^{\,4}\big)+{9\over 2}t_0 t_1 t_2\big[5+4x_2
+2x_0x_1(1+2 x_2)\big]\nonumber\\ && \times \big(\vec l_{12}^{\,\,2}\,\vec 
l_{13}^{\,\,2}-\vec q^{\,4}\big)+{9\over 4}t_0 t_2^2(5+8x_2+2x_2^2) \big(\vec 
l_{12}^{\,\,2}\,\vec l_{13}^{\,\,2}-2\vec l_{12}^{\,\,2} \,\vec q^{\,2}+\vec q^{\,4}
\big)\nonumber\\ &&+{9\over 16}t_1^2 t_2\big[5+2x_1^2+
4x_2(1+x_1^2)\big] \big(\vec l_{12}^{\,\,2}\,\vec l_{13}^{\,\,2}\,\vec l_{23}^{\,\,2}
+\vec l_{12}^{\,\,2}\,\vec l_{13}^{\,\,2}\,\vec q^{\,2}-\vec l_{12}^{\,\,2}\,
\vec q^{\,4}-\vec q^{\,6} \big)\nonumber\\ &&+{9\over 16}t_1 t_2^2(5+8x_2+2x_2^2)
\big(\vec l_{12}^{\,\,2}\,\vec l_{13}^{\,\,2}\,\vec l_{23}^{\,\,2}- \vec l_{12}^{\,\,2}
\,\vec l_{13}^{\,\,2}\,\vec q^{\,2}-\vec l_{12}^{\,\,2}\,\vec q^{\,4}+\vec q^{\,6}
\big)\nonumber\\ &&+{3\over 16}t_1^3(1-6x_1^2)\big(\vec l_{12}^{\,\,2}\,\vec 
l_{13}^{\,\,2}\,\vec l_{23}^{\,\,2}+3 \vec l_{12}^{\,\,2}\,\vec l_{13}^{\,\,2}\,
\vec q^{\,2}+3\vec l_{12}^{\,\,2}\,\vec q^{\,4}+\vec q^{\,6}\big)\nonumber\\ &&
+{t_2^3\over 16}(35+84x_2+78x_2^2+28x_2^3)\big(\vec l_{12}^{\,\,2}\,\vec 
l_{13}^{\,\,2}\,\vec l_{23}^{\,\,2}-3 \vec l_{12}^{\,\,2}\,\vec l_{13}^{\,\,2}\,
\vec q^{\,2}+3\vec l_{12}^{\,\,2}\,\vec q^{\,4}-\vec q^{\,6}\big)\nonumber\\ && 
+9W_0^2(\vec l_{12}\!\times\! \vec q\,)\!\cdot\! (\vec l_{13}\!\times\! \vec q
\,) \big\{4t_0(1+x_0)+t_1(1+x_1)(\vec l_{23}^{\,\,2}+\vec q^{\,2})+5t_2(1+x_2)
(\vec l_{23}^{\,\,2}-\vec q^{\,2})\big\}\,.\end{eqnarray}
Here, $\vec l_{ij} = \vec l_i -\vec l_j$ denote differences of the loop-momenta 
$\vec l_1,  \vec l_2,\vec l_3$ belonging to individual nucleon rings and 
$\vec q$ is the momentum running  into and out of each ring (see first diagram in 
Fig.\,1). Note that we have exploited the permutational symmetry in the indices 
$(123)$ in order to reduce the number of independent terms. It is interesting to 
observe that, as a result of the triple spin-trace, the spin-orbit interaction 
$\sim W_0$ contributes only at quadratic order. 

The next step in the evaluation of the energy density consists in performing four 
four-dimensional loop-integrations. The loop-integrations related to the three 
closed nucleon-rings can actually be factorized (by means of tensor 
contractions), because the interaction product in eq.(4) involves only products of 
scalar-products of the momenta $\vec l_1,  \vec l_2,\vec l_3$ and $\vec q$. By 
exploiting this convenient factorization, the three internal (four-dimensional)
loop-integrations can be solved analytically in terms of cubic expressions in a 
set of (euclidean) polarization functions.\footnote{The advantage of the (basic) 
euclidean polarization function is that it is real-valued, whereas the one in 
Minkowski-space is complex-valued (see ref.\cite{fetter}).} 

\subsection{Polarization functions}
An elementary way to derive the euclidean polarization function is to start from 
the finite-temperature formalism, with the baryon chemical potential set to $\mu= 
k_f^2/2M$, and to take the limit $T\to 0$ in the end. The sum over 
(internal) fermionic Matsubara frequencies leads to Fermi-Dirac distributions, 
which degenerate to step-functions in the limit $T\to 0$, while the external 
bosonic Matsubara frequency becomes a continuous variable $\omega$. Carrying out 
this procedure, the euclidean polarization function including a minus-sign from the 
closed fermion loop has the representation:
\begin{equation} \Pi(\omega,\vec q\,) = \int\!{d^3 l \over (2\pi)^3}\,{1\over 
i \omega+\vec l \cdot \vec q/M} \Big\{\theta(k_f-|\vec l-\vec q/2|)-\theta(
k_f-|\vec l+\vec q/2|) \Big\} \,.\end{equation}
The integral in eq.(5) is most conveniently solved by substituting $\vec l \to 
-\vec l$ in the second term and then shifting $\vec l \to \vec l+ \vec q/2$. 
This way the difference of the two emerging energy denominators becomes 
real-valued. It is furthermore advantageous to introduce dimensionless 
variables $s$ and $\kappa$ by setting $|\vec q\,| = 2s k_f$ and $\omega = 2 s 
\kappa\, k_f^2/M$. Altogether one finds:
\begin{equation} \Pi[1](\omega,\vec q\,)={Mk_f \over4\pi^2 s}\,Q_0(s,\kappa)
\,,\end{equation} 
with the dimensionless function:
\begin{equation}Q_0(s,\kappa) = s - s\kappa \arctan{1+s \over \kappa} - 
s\kappa \arctan{1-s \over \kappa}+{1\over 4}(1-s^2+\kappa^2) \ln{(1+s)^2+
\kappa^2 \over (1-s)^2+\kappa^2}\,.\end{equation}
It is important to note that this expression for $Q_0(s,\kappa)$ agrees perfectly 
with ref.\cite{runge}, where a completely different method has been used to derive 
the polarization function. For the purpose of factorizing products of 
scalar-products (of $\vec l_1, \vec l_2,\vec l_3$ and $\vec q\,$) one needs 
additional polarization functions, which involve one or two $\vec l$-factors:
\begin{equation} \Pi[\vec l\,]=- {Mk_f^2 \over 4\pi^2 s}\, i\kappa\, Q_0(s,
\kappa)\, \hat q\,,\qquad \Pi[l_il_j]= {Mk_f^3\over 4\pi^2 s}\bigg\{
{\delta_{ij} \over 3}\,  Q_1(s,\kappa)+\Big( \hat q_i \hat q_j -{\delta_{ij} 
\over 3} \Big)\, Q_2(s,\kappa)\bigg\} \,, \end{equation} 
where
\begin{equation}Q_1(s,\kappa)=(1-s^2)Q_0(s,\kappa)+{s\over 2}(1+s^2+\kappa^2) 
-{1\over 8}\big[(1+s)^2+\kappa^2\big]\!\big[(1-s)^2+\kappa^2\big]\ln {(1+s)^2
+\kappa^2 \over (1-s)^2+\kappa^2}\,, \end{equation} 
\begin{equation}Q_2(s,\kappa)= s-{1\over 2} Q_1(s,\kappa)-{3 \kappa^2 \over 2}
Q_0(s,\kappa)\,, \end{equation}
as well as those involving three or four $\vec l$-factors combined to 
a vector or a scalar:
\begin{equation} \Pi[\vec l^{\,2}\vec l\,]=- {Mk_f^4 \over 4\pi^2 s}\,i\kappa
\, Q_1(s,\kappa)\, \hat q\,,\qquad \Pi[\vec l^{\,4}]= {Mk_f^5\over 4\pi^2 s}
\, Q_3(s,\kappa)\,, \end{equation} 
where
\begin{equation}Q_3(s,\kappa)={16 s^3 \over 9} + {2\over 3}(2-2s^2-\kappa^2)
Q_1(s,\kappa)+{1\over 3}\big[2\kappa^2(1-3s^2)-(s^2-1)^2\big]Q_0(s,\kappa)\,. 
\end{equation} 
It is worth to note that all dimensionless polarization functions 
$Q_j(s,\kappa)$ are even under $\kappa \to -\kappa$. 
\section{Energy per particle}
In the previous section we have assembled all the  necessary ingredients to 
evaluate the particle-hole ring diagrams at four-loop order. The remaining 
four-dimensional loop-integral $(2\pi)^{-4}\!\int_{-\infty}^\infty d\omega\!\int\!d^3 
q$ can be reduced to a double-integral of the form $(8k_f^5/M\pi^3)\int_0^\infty \!
d\kappa \int_0^\infty\!ds\, s^3$. By adapting the pattern of cubic terms in the 
Skyrme parameters $t_j,x_j,W_0$ as specified by the interaction product in eq.(4), 
one arrives at the following expression for the three-ring energy per particle of 
isospin-symmetric nuclear matter (with density $\rho = 2k_f^3/3\pi^2$):
\begin{eqnarray} \bar E(k_f)^{\rm (3-ring)} &=& {M^2 k_f^5 \over 32 \pi^7} \bigg\{ 
t_0^3(1-6x_0^2){\cal N}_1 + k_f^2 \,t_0^2 t_1(1-2x_0^2-4x_0 x_1){\cal N}_2
\nonumber \\ &&+ k_f^2 \, t_0^2 t_2\big[5+4x_2+2x_0^2(1+2x_2)\big]{\cal N}_3  
+k_f^4 \, t_0 t_1^2(4x_0x_1+2x_1^2-1){\cal N}_4\nonumber \\ && +k_f^4\,t_0t_1t_2
\Big[{5\over 2}+x_0x_1(1+2x_2)+2x_2\Big] {\cal N}_5  +k_f^4 \, t_0 t_2^2
\Big[{5\over 2}+4x_2+x_2^2\Big] {\cal N}_6 \nonumber \\ &&+k_f^6 \, t_1^2 t_2
\Big[{5\over 2}+x_1^2+2x_2(1+x_1^2)\Big] {\cal N}_7 +k_f^6 \, t_1 t_2^2
\Big[{5\over 2}+4x_2+x_2^2\Big] {\cal N}_8\nonumber \\ &&+k_f^6\,t_1^3(1-6x_1^2)
{\cal N}_9+k_f^6\, t_2^3\Big[{5\over 4}+3x_2+{39 \over 14}x_2^2+x_2^3\Big] 
{\cal N}_{10} \nonumber \\ && +k_f^4 \,W_0^2\Big[t_0(1+x_0) {\cal N}_{11} +k_f^2 
\, t_1(1+x_1) {\cal N}_{12}  +k_f^2 \,  t_2(1+x_2) {\cal N}_{13}\Big]\bigg\} 
\,. \end{eqnarray}
The proportionality of $\bar E(k_f)^{\rm (3-ring)}$ to the squared nucleon mass $M^2$ 
is obvious for a third-order contribution and the occurring odd powers of the Fermi 
momentum $k_f$ are predetermined by the dimensions of the coupling 
parameters: $t_0 \sim \text{fm}^2$ and $ t_{1,2}, W_0 \sim \text{fm}^4$. Beyond 
these scaling properties, 
the entire many-body dynamics as represented by the third-order particle-hole 
ring diagrams is encoded 
in the numerical coefficients ${\cal N}_j,\, j =1, \dots, 13$, which will be 
computed in the next subsection.  For comparison, we consider also the 
third-order 
particle-hole contribution in pure neutron matter. In this case the 
isospin-exchange operator $P_\tau\to 1$ and the antisymmetrized Skyrme 
$nn$-contact interaction depends only on the parameter combinations $t_0(1-x_0), 
t_1(1-x_1), t_2(1+x_2)$ and $W_0$. Taking care of these modifications, one 
finds for the three-ring energy per particle in pure neutron matter (with density 
$\rho_n = k_n^3/3\pi^2$):
\begin{eqnarray} \bar E_n(k_n)^{\rm (3-ring)} &=& {M^2 k_n^5\over 96 \pi^7} \bigg\{ 
t_0^3(x_0-1)^3 {\cal N}_1 + k_n^2 \,t_0^2 t_1(x_0-1)^2(x_1-1){\cal N}_2
\nonumber \\ && + k_n^2 \, t_0^2 t_2(x_0-1)^2(x_2+1)3{\cal N}_3  
+k_n^4 \, t_0 t_1^2(1-x_0)(x_1-1)^2{\cal N}_4  \nonumber \\ && +k_n^4\,t_0t_1t_2
(x_0-1)(x_1-1)(x_2+1){3{\cal N}_5\over 2}  +k_n^4 \, t_0 t_2^2(1-x_0)
(1+x_2)^2\, {3{\cal N}_6 \over 2}\nonumber \\ &&+k_n^6 \, t_1^2 t_2(1-x_1)^2
(1+x_2){3{\cal N}_7\over 2} +k_n^6 \, t_1 t_2^2(1-x_1)(1+x_2)^2\, {3{\cal N}_8
\over 2}\nonumber \\ &&+k_n^6\,t_1^3(x_1-1)^3 {\cal N}_9+k_n^6\, t_2^3(1+x_2)^3
\,{45{\cal N}_{10} \over 28}+k_n^6 \, W_0^2 t_2(1+x_2) {8 {\cal N}_{13} \over 5}
\bigg\} \,, \end{eqnarray} 
which differs from eq.(13) by certain reduction factors at the coefficients
${\cal N}_j$. Evidently, the terms proportional to $W_0^2t_0(1+x_0) {\cal N}_{11}$ 
and $W_0^2t_1(1+x_1) {\cal N}_{12}$ have dropped out in $\bar E_n(k_n)^{\rm (3-ring)}$ 
since these involve a coupling to the Pauli-forbidden $^3S_1$-state of two 
neutrons.    
\subsection{Calculation of four-loop coefficients}
We are left with the task to compute numerically the four-loop coefficients 
${\cal N}_j$. In the case of the pure s-wave contact-interaction $t_0(1+x_0P_\sigma)$ 
this concerns ${\cal N}_1$, which is calculated most efficiently by  introducing
polar coordinates, $ s = r \cos\varphi,\, \kappa = r \sin\varphi$, in the 
$s\kappa$ quarter-plane:
\begin{equation} {\cal N}_1 = 12\int\limits_0^\infty \!\!dr\!\!\int
\limits_0^{\pi/2}\!\!d\varphi \,r\,[Q_0(s,\kappa)]^3 = 4.1925784\,.\end{equation}
At this point one can remark that the coefficient in the left part of eq.(1) is 
$2{\cal N}_1/3 =2.7950523$.  The s-wave effective range correction $\sim t_1(1+x_1
P_\sigma)$ is involved linearly in 
the second terms in eqs.(13,14) and the corresponding coefficient ${\cal N}_2$ is 
determined by an integral over $Q_0^2\big[Q_1+(2s^2+\kappa^2)Q_0\big]$, which 
however diverges. We apply dimensional regularization in a practical and empirical
way by subtracting those terms from the integrand which lead to power divergencies 
$r_{\text{max}}^{2n+1}$ in the radial integration $\int_0^{r_\text{max}}\!dr$ up to a cutoff 
$r_{\text{max}}$. After implementing this regularization the (convergent) coefficient 
${\cal N}_2$ is given by: 
\begin{equation} {\cal N}_2 = \int\limits_0^\infty \!\!dr\!\!\int
\limits_0^{\pi/2}\!\!d\varphi \Big\{18 rQ_0^2\big[Q_1+(2s^2+\kappa^2)Q_0\big]-{
16\over 3} \cos^3\!\varphi\,(2+\cos 2\varphi)\Big\}=-0.4633512\,.\end{equation}
In the actual calculation it is most advantageous to expand the integrand further 
in powers of $r^{-2}$ up to order $r^{-4}$. The value 
$\int_0^{\pi/2}\!d\varphi\big\{f_2(\cos\varphi)/r_{\rm max}+f_4(\cos\varphi)
/3r^3_{\rm max}\big\}$ is then included as a good approximation of the contribution 
to the double-integral from the outside region $r>r_{\rm max}$. The angular 
functions $f_{2,4}(\cos\varphi)$ emerging from the $1/r^2$-expansion are always odd 
polynomials in $\cos\varphi$. By applying this procedure one gets numerically 
accurate and well-converged results from the restricted radial integral 
$\int_0^{r_\text{max}}dr$ with $r_{\rm max}$ in the range $20$ to $50$. The regularized 
coefficient ${\cal N}_3$ for the ssp-wave interference term proportional to $t_0^2 
t_2$ in eqs.(13,14) reads: 
\begin{equation} {\cal N}_3 = \int\limits_0^\infty \!\!dr\!\!\int
\limits_0^{\pi/2}\!\!d\varphi \Big\{18 rQ_0^2\big[Q_1+(\kappa^2-2s^2)Q_0\big]+
{16\over 3} \cos^3\!\varphi\,\cos 2\varphi\Big\} = -2.259163\,.\end{equation}
The product $\vec l_{12}^{\,\,2}\,\vec l_{13}^{\,\,2}$ of two squared 
momentum-differences leads to the combination of polarization functions: 
\begin{equation}{\bf C}_a = 3Q_0Q_1^2+Q_0^2 \Big[Q_3 +{4 \kappa^2 \over 3}
(5Q_1-2Q_2) \Big]\,, \end{equation}
which allows us to compute the next three coefficients:
\begin{equation} {\cal N}_4 = -9\int\limits_0^\infty \!\!dr\!\!\int
\limits_0^{\pi/2}\!\!d\varphi \Big\{{r\over 4}{\bf C}_a +4rs^2Q_0^2\big[Q_1+
(s^2+\kappa^2)Q_0\big]\Big\}_{\rm reg} =  2.902123\,, \end{equation}
\begin{equation} {\cal N}_5 = 9\int\limits_0^\infty \!\!dr\!\!\int
\limits_0^{\pi/2}\!\!d\varphi \Big\{r\big({\bf C}_a -16s^4Q_0^3\big)
\Big\}_{\rm reg} =   2.126584\,, \end{equation}
\begin{equation} {\cal N}_6 = 9\int\limits_0^\infty \!\!dr\!\!\int
\limits_0^{\pi/2}\!\!d\varphi \Big\{{r\over 2}{\bf C}_a +8rs^2Q_0^2\big[
(s^2-\kappa^2)Q_0-Q_1\big]\Big\}_{\rm reg} =  0.438970\,, \end{equation}
where the subscript 'reg' indicates the subtraction of power divergences 
proportional to $r^{2n},\, n=0,1,2$. The product $\vec l_{12}^{\,\,2}\,\vec 
l_{13}^{\,\,2}\,\vec l_{23}^{\,\,2}$ of three squared momentum-differences leads 
to twice the combination:
\begin{equation}{\bf C}_b = \kappa^2Q_0\big(5Q_1^2-8Q_1Q_2+3Q_0Q_3\big)
+3Q_0Q_1Q_3 +{5\over 9} Q_1^3-{8\over 9}Q_2^2(3Q_1+Q_2) \,, \end{equation}
which appears together with ${\bf C}_a$ in the evaluation of the next four 
coefficients:  
\begin{equation} {\cal N}_7 = 9\int\limits_0^\infty \!\!dr\!\!\int
\limits_0^{\pi/2}\!\!d\varphi \Big\{{r\over 4}({\bf C}_b+2s^2{\bf C}_a) -4rs^4
Q_0^2\big[Q_1+(2s^2+\kappa^2)Q_0\big]\Big\}_{\rm reg}=0.48756\,, \end{equation}

\begin{equation} {\cal N}_8 = 9\int\limits_0^\infty \!\!dr\!\!\int
\limits_0^{\pi/2}\!\!d\varphi \Big\{{r\over 4}({\bf C}_b-2s^2{\bf C}_a) +4rs^4
Q_0^2\big[(2s^2-\kappa^2)Q_0-Q_1\big]\Big\}_{\rm reg}=-0.27614\,, \end{equation}

\begin{equation} {\cal N}_9 = 3\int\limits_0^\infty \!\!dr\!\!\int
\limits_0^{\pi/2}\!\!d\varphi \Big\{{r\over 8}({\bf C}_b+6s^2{\bf C}_a) +2rs^4
Q_0^2\big[3Q_1+(2s^2+3\kappa^2)Q_0\big]\Big\}_{\rm reg}=-1.01924\,,\end{equation}

\begin{equation} {\cal N}_{10} = 7\int\limits_0^\infty \!\!dr\!\!\int
\limits_0^{\pi/2}\!\!d\varphi \Big\{{r\over 2}({\bf C}_b-6s^2{\bf C}_a) +8rs^4
Q_0^2\big[3Q_1+(3\kappa^2-2s^2)Q_0\big]\Big\}_{\rm reg}=0.315484\,.\end{equation}
The coefficients belonging to the last three terms in eq.(13) 
involving the spin-orbit coupling $W_0$ at quadratic order, read finally: 
\begin{equation} {\cal N}_{11} = 16\int\limits_0^\infty \!\!dr\!\!\int
\limits_0^{\pi/2}\!\!d\varphi \Big\{6 rs^2Q_0^2(Q_1-Q_2)-{16\over 15} \cos^5\!
\varphi\Big\} = -2.244200\,,\end{equation}
\begin{equation} {\cal N}_{12} = 16\int\limits_0^\infty \!\!dr\!\!\int
\limits_0^{\pi/2}\!\!d\varphi \Big\{rs^2Q_0(Q_1-Q_2)\big[2Q_1+Q_2+3(2s^2+
\kappa^2)Q_0\big]\Big\}_{\rm reg} = -2.30577\,,\end{equation}
\begin{equation} {\cal N}_{13} = 80\int\limits_0^\infty \!\!dr\!\!\int
\limits_0^{\pi/2}\!\!d\varphi \Big\{rs^2Q_0(Q_1-Q_2)\big[2Q_1+Q_2+3(\kappa^2
-2s^2)Q_0\big]\Big\}_{\rm reg} =  2.53887\,.\end{equation}
\subsection{Second order calculation}
In the previous subsection dimensional regularization of the divergent integrals 
${\cal N}_j$ has been performed by subtracting power divergences. 
We will now verify this empirical method by rederiving the known analytical 
results for the Skyrme NN-contact interaction at second order \cite{sk2}. 
The many-body contributions at second order are usually categorized into the 
particle-particle ladder series. But the particle-particle ladder diagrams at 
second order (see Fig.\,1 in ref.\cite{sk2}) can be equally well interpreted 
as the second-order two-ring diagrams of the particle-hole type. When using the 
antisymmetrized NN-interaction one has to include (in addition to symmetry factor 
$1/4$) an extra factor $1/2$ in order not to double-count via $(dir-exc)^2= dir^2
+exc^2-2\,dir\, exc$ the direct (Hartree) and exchange (Fock) term. Performing the 
calculation of the two-ring diagrams with the Skyrme NN-contact interaction, one 
obtains for the energy per particle at second order:  
\begin{eqnarray} \bar E(k_f)^{\rm (2nd)} &=& {3M k_f^4 \over 32 \pi^5} \Big\{ 
t_0^2(1+x_0^2){\cal Z}_1 + k_f^2 \,t_0t_1(1+x_0x_1){\cal Z}_2 \nonumber \\ &&
+ k_f^4 \,t_1^2(1+x_1^2){\cal Z}_3 + k_f^4 \,t_2^2(5+8x_2+5x_2^2){\cal Z}_4
+k_f^4 \,W_0^2{\cal Z}_5\Big\}\,, \end{eqnarray} 
\begin{eqnarray} \bar E_n(k_n)^{\rm (2nd)} &=& {M k_n^4 \over 32 \pi^5} \Big\{ 
t_0^2(1-x_0)^2{\cal Z}_1 + k_n^2 \,t_0t_1(1-x_0)(1-x_1){\cal Z}_2 \nonumber \\ &&
+ k_n^4 \,t_1^2(1-x_1)^2{\cal Z}_3 + k_n^4 \,t_2^2(1+x_2)^2\,9{\cal Z}_4
+k_n^4 \,W_0^2\,2{\cal Z}_5\Big\}\,, \end{eqnarray} 
where the coefficients ${\cal Z}_j$ are given by double-integrals $\int_0^\infty\!d
\kappa \!\int_0^\infty\!ds\,s$ over quadratic expressions in the euclidean 
polarization functions $Q_j(s,\kappa)$. After implementing dimensional 
regularization through the 
subtraction of power divergences, these five coefficients read: 
\begin{equation} {\cal Z}_1 = -8\int\limits_0^\infty \!\!dr\!\!\int
\limits_0^{\pi/2}\!\!d\varphi\Big\{3r s\,Q_0^2-{4\over 3}\cos^3\!\varphi\Big\}
 = 3.451697= {4\pi \over 35}(11-2\ln 2)\,,\end{equation} 
\begin{equation} {\cal Z}_2 = -24\int\limits_0^\infty \!\!dr\!\!\int
\limits_0^{\pi/2}\!\!d\varphi \Big\{r s\,Q_0[Q_1+(2s^2+\kappa^2)Q_0]
\Big\}_{\rm reg} = 3.99902= {8\pi \over 945}(167-24\ln 2)\,,\end{equation} 
\begin{eqnarray} && {\cal Z}_3 = -\int\limits_0^\infty \!\!dr\!\!\int
\limits_0^{\pi/2}\!\!d\varphi \Big\{r s\big[24s^2(s^2+\kappa^2)Q_0^2+3Q_0
\big(4(2s^2+\kappa^2)Q_1+Q_3\big)+5Q_1^2+4Q_2^2\big]\Big\}_{\rm reg} \nonumber \\
&& \qquad  =1.37573= {\pi \over 10395}(4943-564\ln 2)\,,\end{eqnarray} 
\begin{eqnarray} && {\cal Z}_4 = -{1\over 3}\int\limits_0^\infty \!\!dr\!\!\int
\limits_0^{\pi/2}\!\!d\varphi \Big\{r s\big[24s^2(s^2-\kappa^2)Q_0^2+3Q_0
\big(4(\kappa^2-2s^2)Q_1+Q_3\big)+5Q_1^2+4Q_2^2\big]\Big\}_{\rm reg} \nonumber 
\\ && \qquad = 0.0931718 = {\pi \over 31185}(1033-156\ln 2)\,,\end{eqnarray}
\begin{equation} {\cal Z}_5 = 128\int\limits_0^\infty \!\!dr\!\!\int
\limits_0^{\pi/2}\!\!d\varphi \Big\{r s^3Q_0(Q_2-Q_1)\Big\}_{\rm reg} = 
2.70935= {16\pi \over 10395}(631-102\ln 2)\,.\end{equation}
In each case the last entry gives the analytical value of ${\cal Z}_j$ as derived 
in section 3 of refs.\cite{sk2} and one finds perfect agreement with the numerical 
evaluation of the regularized double-integrals. This serves as an important check
on our approach which utilizes the subtraction of power divergences after the 
special choice of polar coordinates $s = r \cos\varphi, \kappa = r \sin\varphi$.    

\section{Tensor interactions}
The Skyrme NN-contact interaction in eq.(2) depends on seven parameters. These are 
two less than entering the general ${\cal O}(p^2)$ NN-contact potential of chiral 
effective field theory \cite{epel,evgeny,machleidt}. A complete matching is 
achieved by adding to $V_{\rm Sk}$ the sum of two Galilei-invariant tensor 
interactions, 
whose antisymmetrized form reads:
\begin{eqnarray} && V_{\rm ten}  -P_\sigma P_\tau V_{\rm ten}\big|_{\vec 
q_{\rm out}\to -\vec q_{\rm out}}= (1-P_\tau)\,t_4\Big\{ \vec \sigma_1\!\cdot\!
\vec q_{\rm out}\,\vec \sigma_2\!\cdot\!\vec q_{\rm out}+\vec \sigma_1\!\cdot\! 
\vec q_{\rm in}\,\vec\sigma_2\!\cdot\!\vec q_{\rm in}-{1\over 3}\vec\sigma_1\!
\cdot\!\vec\sigma_2(\vec q_{\rm out}^{\,2}+ \vec q_{\rm in}^{\,2})\Big\} \nonumber 
\\ && \qquad \qquad\qquad  + (1+P_\tau)\,t_5\Big\{\vec \sigma_1\!\cdot\! \vec 
q_{\rm out}\,\vec \sigma_2 \!\cdot\! \vec q_{\rm in} +\vec \sigma_1\!\cdot\! 
\vec q_{\rm in}\,\vec \sigma_2\!\cdot\! \vec q_{\rm out} - {2\over 3}\vec 
\sigma_1\!\cdot\!\vec \sigma_2 \,\vec q_{\rm out}\!\cdot \vec q_{\rm in}\Big\}\,.
\end{eqnarray}
Note that the spin-exchange operator $P_\sigma$ does not modify the two tensor 
expressions in the curly brackets:  $P_\sigma\{\dots\} =   \{\dots\}$. It is also 
obvious that the $t_4$-term in eq.(37) vanishes in pure neutron matter. The second 
order contributions from $V_{\rm ten}$, when calculated alternatively from the 
two-ring particle-hole diagrams, yield the following expressions for the energy
per particle: 
\begin{equation} \bar E(k_f)^{\rm (2nd)} ={3M k_f^8 \over 32 \pi^5} \big\{ t_4^2 
\,{\cal Z}_6+t_5^2\, {\cal Z}_7\big\} \,, \qquad \bar E_n(k_n)^{\rm (2nd)} ={ M 
k_n^8 \over 16 \pi^5} \,t_5^2\, {\cal Z}_7\,, \end{equation}
with the coefficients ${\cal Z}_6$ and ${\cal Z}_7$ given by:
\begin{eqnarray} &&{\cal Z}_6 = -{16\over 9}\int\limits_0^\infty \!\!dr\!\!\int
\limits_0^{\pi/2}\!\!d\varphi \Big\{r s\big[24s^2(s^2+\kappa^2)Q_0^2+3Q_0
(4\kappa^2Q_1+8s^2Q_2+Q_3)+5Q_1^2+4Q_2^2\big]\Big\}_{\rm reg} \nonumber \\ && 
\qquad = 1.54263 = {32\pi \over 93555}(1525-129\ln 2)\,,\end{eqnarray}
\begin{eqnarray} && {\cal Z}_7= -{16\over 3}\int\limits_0^\infty \!\!dr\!\!\int
\limits_0^{\pi/2}\!\!d\varphi \Big\{r s\big[24s^2(s^2-\kappa^2)Q_0^2+3Q_0
(4\kappa^2Q_1-8s^2Q_2+Q_3)+5Q_1^2+4Q_2^2\big]\Big\}_{\rm reg} \nonumber \\ &&
\qquad =  4.200099 = {32\pi \over 405}(19-3\ln 2)\,.\end{eqnarray}
In both cases one finds perfect agreement between the analytical value derived in section 5 of
ref.\cite{sk2} and the numerical evaluation of a regularized double-integral.

Our next task consists in evaluating the third-order particle-hole ring diagrams
with the tensorial contact-interaction in eq.(37) taking into account also all 
possible interference terms with the Skyrme NN-contact interaction. After some 
tedious algebra related to triple spin-traces one finds the following 
contribution to the three-ring energy per particle of isospin-symmetric nuclear 
matter:   
\begin{eqnarray} \bar E(k_f)^{\rm (3-ring)} &=& {M^2 k_f^9 \over 32 \pi^7} \Big\{ 
k_f^2 \,W_0^2 \big[t_4\,{\cal N}_{14}+t_5\,{\cal N}_{15}\big]\nonumber \\ && + 
t_4^2\big[t_0(x_0-2){\cal N}_{16}+k_f^2\, t_1(x_1-2) {\cal N}_{17}+k_f^2\, t_2
(x_2+2) {\cal N}_{18} \big]\nonumber \\ &&+ t_4 t_5\big[t_0x_0\,{\cal N}_{19}
+k_f^2\, t_1x_1\,{\cal N}_{20}+k_f^2\,t_2x_2\,{\cal N}_{21} \big] \nonumber \\ &&
+ t_5^2\big[t_0(3x_0-2){\cal N}_{22}+k_f^2\, t_1(3x_1-2) {\cal N}_{23}+k_f^2\, 
t_2(3x_2+2) {\cal N}_{24} \big] \nonumber \\ &&+ k_f^2\big[ t_4^3\,{\cal N}_{25}
+t_4^2t_5\,{\cal N}_{26} +t_4 t_5^2\,{\cal N}_{27}+t_5^3\,{\cal N}_{28}\big]\Big\}
\,. \end{eqnarray} 
For pure neutron matter the analogous result is considerably simpler: 
\begin{eqnarray} \bar E_n(k_n)^{\rm (3-ring)} &=& {M^2 k_n^9 \over 24 \pi^7} 
\Big\{ k_n^2 \,{2t_5 \over 5}\big[W_0^2\,{\cal N}_{15}+t_5^2\,{\cal N}_{28}\big]
\nonumber \\ && + t_5^2\big[ t_0(x_0-1) {\cal N}_{22}+ k_n^2\,t_1(x_1-1)
{\cal N}_{23}+k_n^2\,t_2(x_2+1) {\cal N}_{24}\big] \Big\}\,, \end{eqnarray} 
because the tensor term $\sim t_4$ responsible for $^3S_1$-$^3\!D_1$ 
mixing is not operative. We start with the coefficients of the terms in eqs.(41,42)
involving the squared spin-orbit parameter $W_0^2$, which read:
\begin{equation} {\cal N}_{14} = {16\over 3}\int\limits_0^\infty \!\!dr\!\!\int
\limits_0^{\pi/2}\!\!d\varphi \Big\{rs^2Q_0(Q_1-Q_2)\big[5Q_1-11Q_2-6(2s^2+
\kappa^2)Q_0\big]\Big\}_{\rm reg} = 0.872202\,,\end{equation}
\begin{equation} {\cal N}_{15} = {80\over 3}\int\limits_0^\infty \!\!dr\!\!\int
\limits_0^{\pi/2}\!\!d\varphi \Big\{rs^2Q_0(Q_1-Q_2)\big[5Q_1-11Q_2+6(2s^2-
\kappa^2)Q_0\big]\Big\}_{\rm reg} = -5.01745\,.\end{equation}
In the process of factorizing the occurring products of scalar-products\,\footnote{
For example: $(\vec l_1\!\cdot\!\vec l_2)^3$ leads to the cubic expression 
$\kappa^2 Q_0(6Q_1 Q_5-3Q_1^2-5Q_5^2)/2$, and $(\vec l_1\!\cdot\!\vec l_2)
(\vec l_1\!\cdot\!\vec l_3)(\vec l_2\!\cdot\!\vec l_3)$ leads to $(Q_1^3+6Q_1
Q_2^2+2Q_2^3)/9$.} (of $\vec l_1, \vec l_2,\vec l_3$  and $\vec q$\,) one 
encounters new polarization functions:
\begin{equation} \Pi[\vec l^{\,2} (\vec l \!\cdot\!\hat q)^2] = {M k_f^5 \over 
4\pi^2 s}\, Q_4(s,\kappa)\,, \qquad \Pi[(\vec l \!\cdot\!\hat q)^3]=-{M k_f^4 
\over 4\pi^2 s}\, i\kappa\, Q_5(s,\kappa)\,, \end{equation}
which introduce the dimensionless functions: 
\begin{equation} Q_4(s,\kappa)= {2s \over 3}(1+s^2)- \kappa^2  Q_1(s,\kappa)
\,, \qquad Q_5(s,\kappa)={2s\over 3}-\kappa^2 Q_0(s,\kappa)\,,\end{equation}
where $Q_1(s,\kappa)$ is written in eq.(9) and $Q_0(s,\kappa)$ in eq.(7).
The interference terms quadratic in $t_4,\,t_5$ lead to the following three
combinations of the dimensionless polarization functions:
\begin{equation}{\bf K}_a = 2Q_0^2\Big[ Q_3+{2\kappa^2\over 3}(Q_1+5Q_2)\Big] 
+6Q_0 Q_2^2\,,\end{equation}
\begin{equation}{\bf K}_b = 8s^2\Big\{Q_0^2\Big[ 3Q_4-Q_3
+{2\kappa^2\over 3}(5Q_1-2Q_2+9Q_5)\Big]+6Q_0Q_1Q_2\Big\}\,,\end{equation}
\begin{eqnarray}{\bf K}_c &=& Q_0\big[4\big(Q_1 Q_3-Q_2 Q_3+3Q_2 Q_4\big)
+\kappa^2\big(4Q_1 Q_2-Q_1^2-18Q_1 Q_5-12Q_2 Q_5+15Q_5^2\big)\big]\nonumber 
\\ &&+12\kappa^2Q_0^2Q_4-{4\over 9}\big(10Q_1^3+15Q_1 Q_2^2+2Q_2^3\big)\,, 
\end{eqnarray} 
which allow to compute the next sequence of nine coefficients ${\cal N}_j$:  
\begin{equation} {\cal N}_{16} = 2\int\limits_0^\infty \!\!dr\!\!\int
\limits_0^{\pi/2}\!\!d\varphi \Big\{r {\bf K}_a+32rs^2Q_0^2\big[Q_2+(s^2+
\kappa^2)Q_0\big]\Big\}_{\rm reg} = -2.91600\,,\end{equation}
\begin{equation} {\cal N}_{17} = \int\limits_0^\infty \!\!dr\!\!\int
\limits_0^{\pi/2}\!\!d\varphi \Big\{{r\over 2} \big(4s^2{\bf K}_a+{\bf K}_b+{\bf K}_c
\big)+32r s^4Q_0^2 \big[Q_1+2Q_2 +(2s^2+3\kappa^2) Q_0\big]\Big\}_{\rm reg} = -2.78344
\,,\end{equation}
\begin{equation} {\cal N}_{18} = \int\limits_0^\infty \!\!dr\!\!\int
\limits_0^{\pi/2}\!\!d\varphi \Big\{{r\over 2} \big({\bf K}_b+{\bf K}_c-4s^2{\bf K}_a
\big)+32r s^4Q_0^2 \big[Q_1-2Q_2 -(2s^2+\kappa^2) Q_0\big]\Big\}_{\rm reg} = 
0.422018\,,\end{equation}
\begin{equation} {\cal N}_{19} = 12\int\limits_0^\infty \!\!dr\!\!\int
\limits_0^{\pi/2}\!\!d\varphi \Big\{r \big({\bf K}_a-32s^4Q_0^3\big)
\Big\}_{\rm reg} = 10.1541\,,\end{equation}
\begin{equation} {\cal N}_{20} = 3\int\limits_0^\infty \!\!dr\!\!\int
\limits_0^{\pi/2}\!\!d\varphi \Big\{r (4s^2{\bf K}_a+{\bf K}_c)-64r s^4Q_0^2 
\big[Q_1+(2s^2+\kappa^2) Q_0\big]\Big\}_{\rm reg} = 8.05637\,,\end{equation}
\begin{equation} {\cal N}_{21} = 3\int\limits_0^\infty \!\!dr\!\!\int
\limits_0^{\pi/2}\!\!d\varphi \Big\{r ({\bf K}_c-4s^2{\bf K}_a)+64r s^4Q_0^2 
\big[(2s^2-\kappa^2) Q_0-Q_1\big]\Big\}_{\rm reg} = -0.745696\,,\end{equation}
\begin{equation} {\cal N}_{22} = 6\int\limits_0^\infty \!\!dr\!\!\int
\limits_0^{\pi/2}\!\!d\varphi \Big\{r{\bf K}_a+32rs^2Q_0^2\big[(s^2-\kappa^2)Q_0
-Q_2\big]\Big\}_{\rm reg} = -1.07623\,,\end{equation}
\begin{equation} {\cal N}_{23} = 3\int\limits_0^\infty \!\!dr\!\!\int
\limits_0^{\pi/2}\!\!d\varphi \Big\{{r\over 2} \big(4s^2{\bf K}_a-{\bf K}_b+
{\bf K}_c\big)+32r s^4Q_0^2\big[Q_1-2Q_2+(2s^2-\kappa^2)Q_0\big]\Big\}_{\rm reg} 
= -3.23779\,,\end{equation}
\begin{equation} {\cal N}_{24} = 3\int\limits_0^\infty \!\!dr\!\!\int
\limits_0^{\pi/2}\!\!d\varphi \Big\{{r\over 2} \big({\bf K}_c-{\bf K}_b-4s^2
{\bf K}_a\big)+32r s^4Q_0^2\big[Q_1+2Q_2+(3\kappa^2-2s^2) Q_0\big]\Big\}_{\rm reg} 
= 1.30198\,.\end{equation}
Finally, the contributions cubic in the tensor-couplings $t_4,\,t_5$ lead to the  
combinations:
\begin{equation} {\bf K}_d= 4s^2Q_0^2\big[3Q_3-9Q_4+2\kappa^2(4Q_1-7Q_2-9Q_5)
\big]-72s^2 Q_0 Q_2^2\,,\end{equation}
and 
\begin{eqnarray}{\bf K}_e &=& 6Q_0\big(Q_2+\kappa^2 Q_0\big)\big(Q_3-3Q_4\big)
+{2\over 9}\big(35Q_1^3+21Q_1 Q_2^2-2Q_2^3\big)\nonumber \\ && +\kappa^2Q_0\Big(
27Q_1 Q_5-22Q_1 Q_2+18Q_2 Q_5+{23 \over 2}Q_1^2-{45\over 2}Q_5^2\Big)\,, 
\end{eqnarray} 
of the euclidean polarization functions which allow us to compute the last four 
coefficients ${\cal N}_j$:
\begin{equation} {\cal N}_{25} ={2\over 9}\int\limits_0^\infty \!\!dr\!\!\int
\limits_0^{\pi/2}\!\!d\varphi \Big\{r ({\bf K}_d+{\bf K}_e)
-64r s^4Q_0^2\big[3Q_2+(2s^2+3\kappa^2)Q_0 \big]\Big\}_{\rm reg} = 0.758472
\,,\end{equation}

\begin{equation} {\cal N}_{26} =-2\int\limits_0^\infty \!\!dr\!\!\int
\limits_0^{\pi/2}\!\!d\varphi \Big\{{r\over 3} ({\bf K}_d+3{\bf K}_e)
+64r s^4Q_0^2\big[Q_2+(2s^2+\kappa^2)Q_0 \big]\Big\}_{\rm reg} = 5.01569
\,,\end{equation}
\begin{equation} {\cal N}_{27} =2\int\limits_0^\infty \!\!dr\!\!\int
\limits_0^{\pi/2}\!\!d\varphi \Big\{{r\over 3} ({\bf K}_d -3{\bf K}_e)
+64r s^4Q_0^2\big[(2s^2-\kappa^2) Q_0-Q_2\big]\Big\}_{\rm reg} = -1.83226
\,,\end{equation}
\begin{equation} {\cal N}_{28} = {10\over 3}\int\limits_0^\infty \!\!dr\!\!\int
\limits_0^{\pi/2}\!\!d\varphi \Big\{r ({\bf K}_d-{\bf K}_e)
+64r s^4Q_0^2\big[3Q_2+(3\kappa^2-2s^2) Q_0\big]\Big\}_{\rm reg} = 5.40146
\,.\end{equation}
One observes that the entire set of the 28 ${\cal N}_j$-coefficients spans in 
magnitude a range from $0.28$ up to $10$ and several subgroups are similar in 
size. In view of the complexity of a four-loop calculation, it is remarkable that 
all coefficients ${\cal N}_j$ could be computed with such a high numerical 
precision.

\section{Third-order ring diagrams with one-pion exchange}
In this section we evaluate the third-order particle-hole ring diagrams shown in 
Fig.\,1 with the longest-range component of the NN-interaction, namely the one-pion 
exchange. After antisymmetrization the $1\pi$-exchange potential in momentum space
takes the (Galilei-invariant) form:
\begin{eqnarray} V_{1\pi} &=& -{g_A^2 \over 4f_\pi^2} \bigg\{ \vec \tau_1\!\cdot\!
\vec\tau_2 \, {\vec \sigma_1\!\cdot\!(\vec q_{\rm in}\!-\! \vec q_{\rm out})\,\vec 
\sigma_2\!\cdot\!(\vec q_{\rm in}\!-\!\vec q_{\rm out}) \over m_\pi^2+(\vec 
q_{\rm in}\!-\!\vec q_{\rm out})^2} \nonumber \\ && +{1 \over 4}(\vec \tau_1\!
\cdot\!\vec\tau_2-3)\,{2\vec \sigma_1\!\cdot\!(\vec q_{\rm in}\!+\!\vec q_{\rm out})
\,\vec \sigma_2\!\cdot\!(\vec q_{\rm in}\!+\!\vec q_{\rm out}) +(1-\vec \sigma_1\!
\cdot\!\vec\sigma_2)(\vec q_{\rm in}\!+\!\vec q_{\rm out})^2\over m_\pi^2+(\vec 
q_{\rm in}\!+\!\vec q_{\rm out})^2} \bigg\}\,,\end{eqnarray}
with $g_A=1.29$ the nucleon axial-vector coupling constant, $f_\pi = 
92.2\,$MeV the pion decay constant, and $m_\pi = 135\,$MeV the (neutral) pion 
mass. The direct diagram $I=dir^3$ is very easily evaluated since the interaction 
product depends only on the momentum $\vec q$ flowing through the three
polarization bubbles.  The pertinent contribution to the energy per particle of 
isospin-symmetric nuclear matter reads:   
\begin{equation}\bar E(k_f)^{\rm (I)} = - {3g_A^6 M^2 k_f^5 \over 32 \pi^7 f_\pi^6} 
\int\limits_0^\infty\!ds \!\int\limits_0^\infty\!d\kappa \bigg[{s^2 Q_0(s,\kappa) 
\over s^2+\beta}\bigg]^3\,, \end{equation}
with the ratio $\beta = m_\pi^2/4k_f^2$. Note that the expression for 
$\bar E(k_f)^{\rm (I)}$ could also be inferred from the photon-exchange ring energy 
discussed in textbooks \cite{fetter,runge}. For the exchange-type 
diagram  $II=-dir^2\, exc$ only one internal loop-integral can solved in terms 
of the polarization function $\Pi(\omega, \vec q\,)$ introduced in eq.(5). 
Putting all the pieces together, one ends up with the following representation of 
the energy per particle:
\begin{eqnarray} \bar E(k_f)^{\rm (II)} &=&  {9g_A^6 M^2 k_f^5 \over (2\pi)^7 
f_\pi^6}\int\limits_0^\infty\!\!ds \!\!\int\limits_0^\infty\!\!d\kappa\!\!\int
\limits_0^1\!\!dl_1 \!\!\int\limits_0^1\!\!dl_2\!\!\int\limits_{-l_1}^{l_1}\!\!dx 
\!\!\int\limits_{-l_2}^{l_2}\!\!dy \, {l_1l_2s^4\,Q_0(s,\kappa)\,(s^2+\beta)^{-2}
\over [(s+x)^2+\kappa^2][(s+ y)^2+\kappa^2]}\nonumber \\ && \times \bigg\{(s+ x)
(s+y) \nonumber+\big[2\beta+(x-y)^2\big] \big[\kappa^2-(s+x)(s+y)\big] 
W_a^{-1/2}\ \nonumber \\ && -\big[2\beta+(2s+x+y)^2\big] \big[\kappa^2+(s+x)(s+y)
\big]W_b^{-1/2}\bigg\}\,,\end{eqnarray}
with the radicands $W_a= \big[4\beta+l_1^2+l_2^2-2x y\big]^2-4(l_1^2-x^2)(l_2^2-y^2)$ 
and $W_b=\big[4\beta+l_1^2+l_2^2+4s(s+x+y)+2x y\big]^2-4(l_1^2-x^2)(l_2^2-y^2)$. 
Apparently, the factor $s^4\,Q_0(s,\kappa)\,(s^2+\beta)^{-2}$ in the numerator of 
eq.(67) originates from the polarization function $\Pi(\omega, \vec q \,)$ and  
two direct $1\pi$-exchanges depending on $\vec q$. For the remaining two 
exchange-type diagrams $III = dir\,exc^2$ and $IV = - exc^3$ only the 
frequency-integral over the energy denominators $i\omega +\vec l_j\cdot \vec q/M$ 
from the polarization functions (see eq.(5)) can be performed analytically and 
residue calculus provides the relevant formula:
\begin{equation}\int\limits_{-\infty}^\infty \! {d\omega \over 2\pi} {1  
\over (i \omega+a)(i \omega+b)(i \omega+c)} = {\theta(a) \over (a-b)(a-c)} +  
{\theta(b) \over (b-a)(b-c)} + {\theta(c) \over (c-a)(c-b)}\,,  \end{equation}
with $a, b, c$ three real-valued parameters. In the end the contributions to the 
energy per particle, $\bar E(k_f)^{\rm (III)}$ and $\bar E(k_f)^{\rm (IV)}$, have 
representations in the form of 9-dimensional integrals over four radii (three 
from $0$ to $1$ and one from $0$ to $\infty$), three directional cosines (from 
$-1$ to $1$) and two azimuthal angles (from $0$ to $2\pi$). The expressions for 
the integrands are very lengthy and not given here. 

In pure neutron matter the four ring diagrams generated by $1\pi$-exchange carry 
different isospin-factors and the energy per particle is composed as:  
\begin{equation}\bar E_n(k_n)={1\over 12}\bar E(k_n)^{\rm (I)}-{1\over 6}\bar 
E(k_n)^{\rm (II)}+{1\over 3}\bar E(k_n)^{\rm (III)}+{1\over 12}\bar E(k_n)^{\rm (IV)}
\,. \end{equation}

\begin{figure}[h!]
\begin{center}
\includegraphics[scale=0.5,clip]{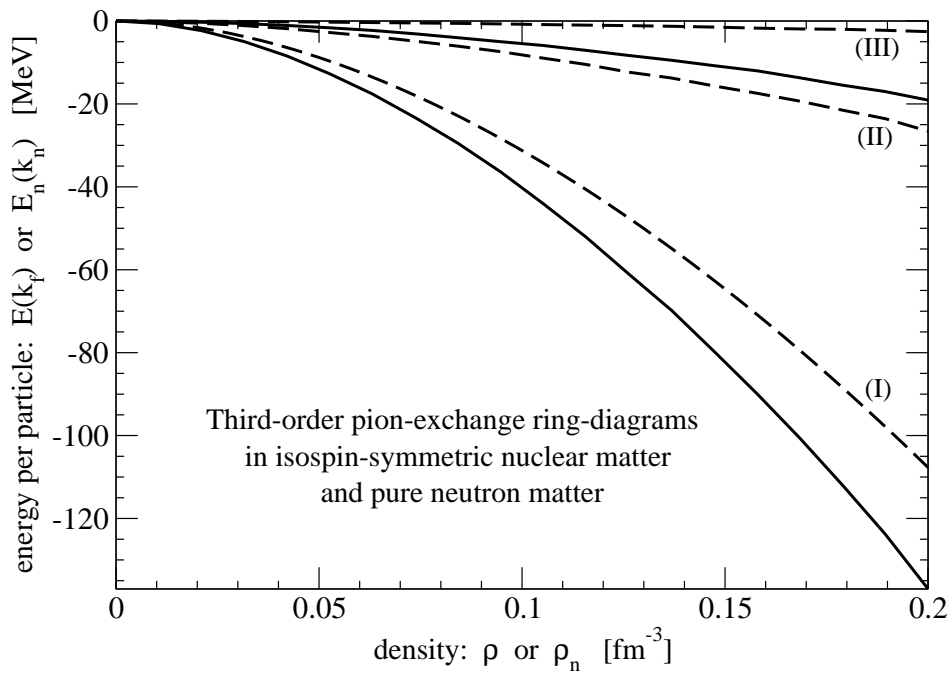}
\end{center}
\vspace{-.6cm}
\caption{Contributions of the third-order particle-hole ring-diagrams generated 
by $1\pi$-exchange to the energy per particle of isospin-symmetric nuclear 
matter and pure neutron matter.}
\end{figure}

Fig.\,2 shows by the dashed lines the contributions from individual ring diagrams 
to the energy per particle $\bar E(k_f)$ as a function of the density $\rho = 
2k_f^3/3\pi^2$. The part $IV=-exc^3$ is so small, that it is not visible on the 
scale of the figure. One observes that the exchange corrections are successively 
suppressed against the direct term, $|III|\ll|II|\ll|I|$, but they are all of the 
same attractive sign. At saturation density $\rho_0=0.16\,$fm$^{-3}$ (or $k_{f0}=
263\,$MeV) one gets from the third-order particle-hole diagrams with long-range 
$1\pi$-exchange a tremendous attraction of $\bar E(k_{f0}) =-92.0\,$MeV. In pure 
neutron matter at the same density it reduces to $\bar E_n(1.26\,k_{f0})=-12.7\,$MeV.
The density-dependence of the pionic three-ring energies is shown by the full 
lines in Fig.\,2. We note as an aside that in the chiral limit $m_\pi = 0$ the 
pionic three-ring energies scale simply with the fifth power of the Fermi momentum
($k_f$ or $k_n$):
\begin{equation}\bar E(k_f)^{3\pi-\rm ring}_{m_\pi = 0} \simeq -{g_A^6M^2 k_f^5 \over
(2\pi f_\pi)^6} \cdot 0.812\,, \qquad\qquad \bar E_n(k_n)^{3\pi-\rm ring}_{m_\pi = 0} 
\simeq - {g_A^6 M^2 k_n^5 \over (4\pi f_\pi)^6} \cdot 2.37\,. \end{equation}
At this point it is important to stress that the results presented in Fig.\,2 are 
in no way representative for the third-order particle-hole contributions computed 
from realistic chiral low-momentum NN-potentials. As one can see from table I in 
ref.\cite{jeremy} the three-ring attraction in isospin-symmetric nuclear matter 
is bounded by about 2\,MeV at $\rho_0=0.16\,$fm$^{-3}$, 
when using chiral NN-potentials with soft cutoffs $\Lambda<500\,$MeV. Nevertheless 
the semi-analytical approach developed here is very helpful in order to benchmark 
the extensive numerical computations \cite{phhole} based on the partial-wave 
decomposition of the NN-potential for test-interactions of the one-boson exchange 
type \cite{jeremy}. Taking for example a pure spin-orbit interaction $\sim i(\vec 
\sigma_1+\vec \sigma_2)\cdot(\vec q_{\rm out}\!\times\!\vec q_{\rm in})$, one finds 
that the third-order ring diagrams vanish identically. In the semi-analytical 
approach this property follows immediately from the triple spin-trace being equal 
to zero, whereas in the partial-wave based method delicate
cancelations in multiple partial-wave sums are at work.

It is also interesting to consider a direct NN-contact interaction $V_{\rm dir} = -C 
\,\vec \tau_1\!\cdot\!\vec\tau_2 \, \vec \sigma_1\!\cdot\!(\vec q_{\rm in}\!-\! 
\vec q_{\rm out})\,\vec \sigma_2\!\cdot\!(\vec q_{\rm in}\!-\!\vec q_{\rm out})$ with 
the same isospin- and spin-dependence as $1\pi$-exchange. Evaluating the four 
diagrams in Fig.\,1 and applying dimensional regularization as done in subsection 
3.1, one finds for the energy per particle: 
\begin{equation}\bar E(k_f) = C^3 M^2 k_f^{11} \pi^{-7}\big(5.73956+0.248478+
0.186314+0 \big) = C^3 M^2 k_f^{11} \pi^{-7}\!\cdot 6.17435\,, \end{equation}
\begin{equation}\bar E_n(k_n)=C^3M^2k_n^{11} \pi^{-7}\!\cdot 0.498988\,.
\end{equation}
It is astonishing that the diagram $IV=-exc^3$ vanishes identically for $V_{\rm dir} 
= -C \,\vec \tau_1\!\cdot\!\vec\tau_2 \, \vec \sigma_1\!\cdot\!(\vec q_{\rm in}\!-\! 
\vec q_{\rm out})\,\vec \sigma_2\!\cdot\!(\vec q_{\rm in}\!-\! \vec q_{\rm out})$. 
In the actual calculation 43 terms had to be translated into cubic expressions in 
$Q_j(s,\kappa)$,  which in the end summed up to zero. This fact gives also an 
explanation for the observed smallness of the 
contribution  $IV=-exc^3$ when evaluated with the long-range $1\pi$-exchange.  
\section{Third-order ladder diagrams}
\begin{figure}[h!]
\begin{center}
\includegraphics[scale=0.6,clip]{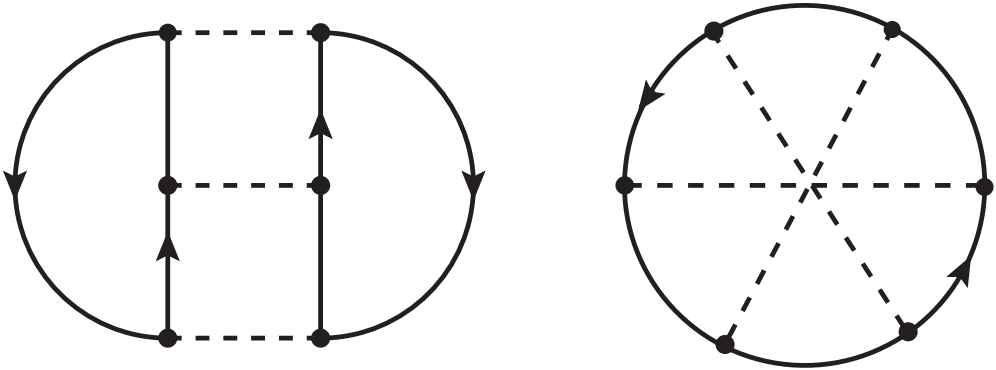}
\end{center}
\vspace{-.6cm}
\caption{Third order ladder diagrams separated into direct and exchange 
contributions.}
\end{figure}
In this section we return to the general ${\cal O}(p^2)$ NN-contact interaction and 
calculate the remaining third-order contributions arising from (particle-particle 
and hole-hole) ladder diagrams. Fig.\,3 shows the direct and exchange-type ladder 
diagrams at third-order with the two-body interaction symbolized by a dashed 
line. We employ the techniques of refs.\cite{resum1,resum2} based on the 
complex-valued in-medium loop, which keeps together particle-particle and 
hole-hole ladder contributions. 

We start with the Skyrme NN-contact interaction $V_{\rm Sk}$ which involves two 
s-wave terms $\sim t_0,t_1$ and two p-wave terms $\sim t_2,W_0$. As shown in 
appendix A of ref.\cite{sk2}, interferences of s-wave and p-wave interactions 
(in ladder diagrams) vanish in a medium with one single Fermi-momentum (see 
also eqs.(5,6) in ref.\cite{resum2}). The case of pure neutron matter is 
particularly simple, because two neutrons interact either in the (odd) $^1S_0$-state
or the (even) $^3P$-states. Taking into account all ($t_0t_1$ and $t_2W_0$) 
interference terms possible at third order, one finds for the energy per particle:  
\begin{eqnarray} \bar E_n(k_n)^{(\rm 3-lad)} &=& {M^2 k_n^5 \over 64\pi^6} \bigg\{
t_0^3(1\!-\!x_0)^3 {\cal B}_1+k_n^2\, t_0^2t_1(1\!-\!x_0)^2(1\!-\!x_1) {\cal B}_2
+k_n^4\, t_0t_1^2(1\!-\!x_0)(1\!-\!x_1)^2 {\cal B}_3 \nonumber \\ && +k_n^6\, 
t_1^3(1\!-\!x_1)^3{\cal B}_4+k_n^6\,t_2^3(1\!+\!x_2)^3 {\cal B}_5+k_n^6\, t_2
(1\!+\!x_2)W_0^2{\cal B}_6+k_n^6\, W_0^3{\cal B}_7\bigg\}\,,\end{eqnarray}
with seven coefficients ${\cal B}_j$ to be determined later. The result for 
isospin-symmetric nuclear matter takes a similar form:
\begin{eqnarray} \bar E(k_f)^{(\rm 3-lad)} &=& {M^2 k_f^5 \over 64\pi^6} \bigg\{
3t_0^3(1\!+\!3x_0^2) {\cal B}_1+3k_f^2\, t_0^2t_1(1\!+\!x_0^2\!+\!2x_0x_1) 
{\cal B}_2  \nonumber \\ && +3k_f^4\, t_0t_1^2(1\!+\!2x_0x_1\!+\!x_1^2){\cal B}_3 
+3k_f^6\, t_1^3(1\!+\!3x_1^2){\cal B}_4 \nonumber \\ &&+k_f^6\,t_2^3\Big({5\over 
3}\!+\!4x_2\!+\!5x_2^2\!+\!{4x_2^3\over 3}\Big) {\cal B}_5+{3k_f^6 \over 2}\, 
t_2(1\!+\!x_2)W_0^2{\cal B}_6+{3k_f^6\over 2}\, W_0^3{\cal B}_7\bigg\}\,,
\end{eqnarray}
where the different dependences on $x_0$ and $x_1$ follow from summing the weights 
in the isotriplet $^1S_0$-state and the isosinglet $^3S_1$-state: $[3(1-x_0)^3+
3(1+x_0)^3]/2 = 3t_0^3(1+3x_0^2)$ and $ [3 (1-x_0)^2(1-x_1)+3 (1+x_0)^2(1+x_1)]/2 = 
3(1+x_0^2+2x_0x_1)$. The factor $1/2$ comes here from dividing energy densities by 
particle densities. For the p-wave term $\sim t_2^3$ the reweighting goes as 
$9(1+x_2)^3\to \big[3(1-x_2)^3+27(1+x_2)^3\big]/2=3(5+12x_2+15x_2^2+4x_2^3)$. For the
interactions $\sim t_2(1+x_2),W_0$ which operate only in isotriplet $^3P$-states, 
one can deduce from a comparison of the energy density in pure neutron matter and 
isospin-symmetric nuclear matter the relation, $3(\text{Hart} + \text{Fock}) = 
4\text{Hart} + 2 \text{Fock}$, between the (double) spin-trace of the direct diagram 
(Hart) and the (single) spin-trace of the exchange-type diagram (Fock). 
Consequently, one has $\text{Hart} =\text{Fock}$ and this explains the relative 
factor $3/2$ for the last two terms in eqs.(73,74).

In order to compute the coefficients ${\cal B}_j$ one needs the in-medium loop,
whose real and imaginary part read \cite{resum1,resum2}:
\begin{equation} R(s,\kappa) = 2 +{1\over 2s}[1-(s+\kappa)^2]\ln{1+s +\kappa
\over |1-s -\kappa|}+{1\over 2s}[1-(s-\kappa)^2]\ln{1+s -\kappa
\over 1-s +\kappa}\,, \end{equation}
\begin{equation} I(s,\kappa) = \kappa \, \theta(1-s-\kappa) + {1\over 2s} 
(1-s^2-\kappa^2)\, \theta(s+\kappa-1)\,.  \end{equation}
Furthermore, iterated p-wave contact-interactions introduce a factor $l_il_j$ and 
the corresponding tensorial in-medium loop is decomposed into a transversal 
projector $\delta_{ij} -\hat P_i \hat P_j$ and a longitudinal projector $\hat P_i 
\hat P_j$, where $\vec P = (\vec p_1+\vec p_2)/2$ with $|\vec p_{1,2}|<k_f$ (for 
details see section 3 in ref.\cite{resum2}). The transversal in-medium loop has 
the following real and imaginary part \cite{resum2}:
\begin{eqnarray} R_\perp(s,\kappa) &=& {2\over 3}-{s^2\over 4}+2\kappa^2+
{(1-\kappa^2)^2\over 4s^2}  +{[(s+\kappa)^2-1]^2\over 16s^3}(s^2+\kappa^2
-4s\kappa-1 )\nonumber \\ && \times \ln{1+s +\kappa \over |1-s -\kappa|} 
+{[(s-\kappa)^2-1]^2\over 16s^3}(s^2+\kappa^2+4s\kappa -1 )\ln{1+s -\kappa 
\over 1-s +\kappa}\,,\end{eqnarray}
\begin{equation} I_\perp(s,\kappa) = \kappa^3 \, \theta(1-s-\kappa) + 
{1-s^2-\kappa^2 \over 16s^3}\Big[12 s^2\kappa^2-(1-s^2-\kappa^2)^2\Big]\,
\theta(s+\kappa-1)\,,  \end{equation}
while the real and imaginary part of the longitudinal in-medium loop read 
\cite{resum2}: 
\begin{eqnarray} R_\parallel(s,\kappa) &=& {8\over 3}+{s^2\over 2}+2\kappa^2
-{(1-\kappa^2)^2\over 2s^2}  +\bigg[{(1-s^2-\kappa^2)^3 \over 8s^3}-\kappa^3
\bigg] \nonumber \\ && \times \ln{1+s +\kappa \over |1-s -\kappa|} +\bigg[
{(1-s^2-\kappa^2)^3 \over 8s^3}+\kappa^3 \bigg] \ln{1+s -\kappa \over 1-s 
+\kappa}\,,\end{eqnarray}
\begin{equation} I_\parallel(s,\kappa) = \kappa^3 \, \theta(1-s-\kappa) + 
{(1-s^2-\kappa^2)^3 \over 8s^3} \,\theta(s+\kappa-1)\,.  \end{equation}
Note that these functions satisfy the relations: $2R_\perp(s,\kappa) +R_\parallel(s,
\kappa) = 4 +3\kappa^2 R(s,\kappa)$ and $2I_\perp(s,\kappa) +I_\parallel(s,\kappa) = 
3\kappa^2 I(s,\kappa)$. 

Now we can specify the formulas for the seven four-loop coefficients ${\cal B}_j, 
j=1,\dots, 7$ together with their numerical values:  
\begin{equation} {\cal B}_1 = 8 \int\limits_0^1ds\,s^2  \int\limits_0^{
\sqrt{1-s^2}}d\kappa\,\kappa\, I(s,\kappa)\big[3R^2(s,\kappa)-\pi^2 I^2(s,
\kappa)\big] = 1.1716223\,,\end{equation} 
which has also been quoted in the right part of eq.(1). The detailed
derivation that $R^2-\pi^2 I^2/3$ is the correct expression for the combined 
third-order ladder contribution has been presented in section 4 of 
ref.\cite{resum1}. From now on we will drop the arguments $s$ and $\kappa$:
\begin{equation} {\cal B}_2 = 8 \int\limits_0^1ds\,s^2  \int\limits_0^{
\sqrt{1-s^2}}d\kappa\,\kappa\, I\big[3\kappa^2(3R^2-\pi^2 I^2)+8R\big] = 
1.9893144\,,\end{equation}
\begin{equation} {\cal B}_3 = 8 \int\limits_0^1ds\,s^2  \int\limits_0^{
\sqrt{1-s^2}}d\kappa\,\kappa\, I\bigg[3\kappa^4(3R^2-\pi^2 I^2)+\bigg(2s^2+14
\kappa^2+{6 \over 5}\bigg) R+4\bigg] = 1.360736\,,\end{equation}
\begin{equation} {\cal B}_4 = 8 \int\limits_0^1ds\,s^2  \int\limits_0^{
\sqrt{1-s^2}}d\kappa\,\kappa\, I\bigg[\kappa^6(3R^2-\pi^2 I^2)+2\kappa^2\bigg(
s^2+3\kappa^2+{3 \over 5}\bigg) R+{4s^2 \over 3}+{8\kappa^2 \over 3}+{4\over 5}
\bigg] = 0.3344923\,,\end{equation}
\begin{equation} {\cal B}_5 = {8\over 9} \int\limits_0^1ds\,s^2  \int\limits_0^{
\sqrt{1-s^2}}d\kappa\,\kappa\, \big[2I_\perp(3R_\perp^2-\pi^2 I_\perp^2)+ 
I_\parallel(3R_\parallel^2-\pi^2 I_\parallel^2)\big] = 0.06699116\,,\end{equation} 
\begin{eqnarray}  {\cal B}_6 &=&  {128\over 9} \int\limits_0^1ds\,s^2  \int
\limits_0^{\sqrt{1-s^2}}d\kappa\,\kappa \Big\{ I_\perp\Big[3R_\perp^2+2R_\perp
R_\parallel+R_\parallel^2 -{\pi^2 \over 3}(3I_\perp^2+2I_\perp I_\parallel+I_\parallel^2) 
\Big] \nonumber \\ &&  +I_\parallel\Big[R_\perp(R_\perp+2 R_\parallel)-{\pi^2 \over 
3}I_\perp(I_\perp +2I_\parallel)\Big]\Big\} = 1.327456\,,
\end{eqnarray} 
\begin{equation}  {\cal B}_7 =  {128\over 9} \int\limits_0^1ds\,s^2  \int
\limits_0^{\sqrt{1-s^2}}d\kappa\,\kappa \Big\{ 2I_\perp\Big(R_\perp R_\parallel -
{\pi^2 \over 3}I_\perp I_\parallel \Big) +I_\parallel \Big(R_\perp^2-{\pi^2 \over 3}
I_\perp^2\Big) \Big\} = 0.4527642\,.\end{equation}
It is worth to note that $I(s,\kappa)$ and $I_{\perp,\parallel}(s,\kappa)$ serve also 
as weight-functions for the reduction of six-dimensional integrals over two Fermi 
spheres $|\vec p_{1,2}|<k_f$ to double-integrals $\int_0^1ds\,s^2  \int_0^{\sqrt{1-s^2}}
\!d\kappa\,\kappa$.
 
Next, we come to the additional third-order ladder contributions generated by the 
tensor contact-interaction $V_{\rm ten}$. One finds for the energy per particle of 
pure neutron matter the result:  
\begin{equation} \bar E_n(k_n)^{(\rm 3-lad)} = {M^2 k_n^{11} \over 64\pi^6} \Big\{
t_5 W_0^2{\cal B}_8+t_5^2t_2(1\!+\!x_2){\cal B}_9+t_5^2 W_0{\cal B}_{10}+t_5^3
{\cal B}_{11}\Big\}\,,\end{equation}
which depends only the new parameter $t_5$. Further p-wave interference terms 
proportional to $t_5 t_2^2$ and $t_5t_2 W_0$ are absent because of 
vanishing spin-traces. Since the tensor term $\sim t_5$ acts only in isovector 
$^3P$-states the expression in eq.(88) reappears in the energy per particle of 
isospin-symmetric nuclear matter: 
\begin{eqnarray} \bar E(k_f)^{(\rm 3-lad)} &=& {M^2 k_f^{11} \over 64\pi^6} \bigg\{
{3\over 2}t_5 W_0^2{\cal B}_8+{3 \over 2}t_5^2t_2(1\!+\!x_2){\cal B}_9+
{3 \over 2}t_5^2 W_0{\cal B}_{10}   \nonumber \\ && +{3 \over 2}t_5^3
{\cal B}_{11} +k_f^{-2} \, t_4^2t_0(1\!+\!x_0){\cal B}_{12}  +t_4^2t_1(1\!+\!x_1)
{\cal B}_{13} \bigg\}\,,\end{eqnarray}
with a factor $3/2$. The last two terms $\sim t_4^2$ in eq.(89) are produced by 
$^3S_1$-$^3\!D_1$ mixing at second order. It has to be combined at third-order 
with an s-wave interaction in the $^3S_1$-channel, which determines the parameter 
combinations $t_{0,1}(1+x_{0,1})$. Note that for third-order interference terms 
all possible orderings need to be considered, since in general these lead to 
different spin-traces.   

Finally, we specify the formulas for the remaining six four-loop coefficients 
${\cal B}_j, j=8,\dots, 13$ together with their numerical values:  
\begin{eqnarray}  {\cal B}_8 &=&  -{256\over 27} \int\limits_0^1ds\,s^2  \int
\limits_0^{\sqrt{1-s^2}}d\kappa\,\kappa \Big\{ I_\perp\Big[3R_\perp^2+11R_\perp
R_\parallel+R_\parallel^2 -{\pi^2 \over 3}(3I_\perp^2+11I_\perp I_\parallel+I_\parallel^2) 
\Big] \nonumber \\ &&  +I_\parallel\Big[R_\perp\Big({11\over 2}R_\perp+2 R_\parallel
\Big)-{\pi^2 \over 6}I_\perp(11I_\perp +4I_\parallel)\Big]\Big\} = -2.243263\,,
\end{eqnarray}
\begin{eqnarray}  {\cal B}_9 &=&  {128\over 9} \int\limits_0^1ds\,s^2  \int
\limits_0^{\sqrt{1-s^2}}d\kappa\,\kappa \Big\{ I_\perp\Big[7R_\perp^2+2R_\perp
R_\parallel+R_\parallel^2 -{\pi^2 \over 3}(7I_\perp^2+2I_\perp I_\parallel+I_\parallel^2) 
\Big] \nonumber \\ &&  +I_\parallel\Big[ R_\perp^2+2R_\perp R_\parallel+2 R_\parallel^2
-{\pi^2 \over 3}(I_\perp^2+2I_\perp I_\parallel +2I_\parallel^2)\Big]\Big\} = 2.042028
\,,\end{eqnarray}
\begin{eqnarray}  {\cal B}_{10} &=&  {256\over 9} \int\limits_0^1ds\,s^2  \int
\limits_0^{\sqrt{1-s^2}}d\kappa\,\kappa \Big\{ I_\perp\Big[6R_\perp^2+7R_\perp
R_\parallel+2R_\parallel^2 -{\pi^2 \over 3}(6I_\perp^2+7I_\perp I_\parallel+2I_\parallel^2
) \Big] \nonumber \\ &&  +I_\parallel\Big[R_\perp\Big({7\over 2}R_\perp+4 R_\parallel
\Big)-{\pi^2 \over 6}I_\perp(7I_\perp +8I_\parallel)\Big]\Big\} = 6.66812\,,
\end{eqnarray}
\begin{eqnarray}  {\cal B}_{11} &=&  -{256\over 81} \int\limits_0^1ds\,s^2  \int
\limits_0^{\sqrt{1-s^2}}d\kappa\,\kappa \Big\{ I_\perp\Big[17R_\perp^2+15R_\perp
R_\parallel+3R_\parallel^2 -\pi^2 \Big({17\over 3}I_\perp^2+5I_\perp I_\parallel+
I_\parallel^2\Big) \Big] \nonumber \\ &&  +I_\parallel\Big[ {15\over 2}R_\perp^2+
6R_\perp R_\parallel+4 R_\parallel^2-\pi^2 \Big({5\over 2}I_\perp^2+2I_\perp I_\parallel 
+{4\over 3}I_\parallel^2\Big)\Big]\Big\} = -1.655323\,,\end{eqnarray}
\begin{eqnarray}  {\cal B}_{12} &=&  32 \int\limits_0^1ds\,s^2  \int
\limits_0^{\sqrt{1-s^2}}d\kappa\,\kappa \Big\{ I_\parallel\Big[R(R_\parallel-R_\perp)
-{\pi^2\over 3} I(I_\parallel-I_\perp) \Big] + I \Big[ 8R\Big( {s^2-2\kappa^2 
\over 3}+{1\over 5}\Big) -{8\over 3} \nonumber \\ &&  + {\kappa^4 \over 2}
(\pi^2 I^2\!-\!3R^2) +R_\perp^2 -{\pi^2 \over 3} I_\perp^2 +{1\over 2}R_\parallel^2 -
{\pi^2 \over 6}I_\parallel^2 +\kappa^2(3R R_\perp\!-\!\pi^2I I_\perp) \Big]\Big\} = 
2.421103\,, \nonumber \\ \end{eqnarray}

\begin{eqnarray}  {\cal B}_{13} &=&  32 \int\limits_0^1ds\,s^2  \int
\limits_0^{\sqrt{1-s^2}}d\kappa\,\kappa \Big\{ I_\parallel\Big[{4\over 9}(3s^2 R+
R_\parallel-R_\perp) +\kappa^2 R(R_\parallel-R_\perp) - \kappa^2 {\pi^2 \over 3} I
(I_\parallel-I_\perp) \Big] \nonumber \\ && +I \Big[ 4\kappa^2\Big( {1\over 5}-
\kappa^2\Big)R  -{8\kappa^2\over 3} +{4\over 3}R_\parallel\Big(s^2+{1\over 5}\Big)
+{4\over 3}R_\perp\Big(\kappa^2+{2\over 5}\Big) + \kappa^2\Big(R_\perp^2 -{\pi^2 
\over 3} I_\perp^2 \Big) \nonumber \\ && +{\kappa^2\over 2}\Big(R_\parallel^2 -
{\pi^2 \over 3}I_\parallel^2\Big) +\kappa^4(3R R_\perp-\pi^2 I I_\perp)  +{\kappa^6 
\over 2}(\pi^2I^2-3R^2) \Big]\Big\} = 1.559127\,.\end{eqnarray}
One observes that most of the four-loop coefficients ${\cal B}_j$ are of similar 
size.
\section*{Appendix: Isospin-asymmetry energy}
In this appendix we evaluate a specific contribution from third-order 
particle-hole ring diagrams to the isospin-asymmetry energy $A(k_f)$. This 
density-dependent quantity is defined by the expansion of the energy per particle of 
isospin-asymmetric nuclear matter, $\bar E_{\rm as}(k_p,k_n) = \bar E(k_f)+ \delta^2 
A(k_f)+{\cal O}(\delta^4)$, with the proton/neutron Fermi momenta set to $k_{p,n} = 
k_f(1\mp \delta)^{1/3}$. The considered two-body interaction is a momentum-independent
NN-contact interaction: 
\begin{equation} V_{\rm ct} =-{\pi \over M} \big[ a_s+3a_t +\vec \sigma_1
\!\cdot\! \vec \sigma_2(a_t-a_s)\big]\,, \end{equation} 
parametrized through the spin-singlet and spin-triplet scattering lengths, 
$a_s=M(x_0-1)/4\pi$ and $a_t = -M(1+x_0)/4\pi$. Since $V_{\rm ct}$ does not change 
the isospin, the evaluation of the four ring diagrams in Fig.\,1 proceeds by 
distinguishing the cases where closed lines represent protons or neutrons. After 
performing the (multiple) spin-traces one obtains from the sum of the ring diagrams 
$I+II+II+IV$ an integrand for the energy density of isospin-asymmetric nuclear 
matter proportional to:
\begin{eqnarray} && {1\over 6} \big[(3a_t+a_s)^3+3(a_t-a_s)^3\big] \big(\Pi_{pp}+
\Pi_{nn}\big)^3 -\big[16a_t^3+(a_s-a_t)^3\big] \big(\Pi_{pp}+\Pi_{nn}\big)\big(
\Pi_{pp}^2+\Pi_{nn}^2\big)  \nonumber \\ && +\big[16a_t^3-(a_s+a_t)^3\big]\big(
\Pi_{pp}^3+\Pi_{nn}^3\big)-{1\over 3}\big[(3a_t+a_s)^3-12a_t^2(3a_s+a_t)\big] \big(
\Pi_{pp}^3+\Pi_{pn}^3+\Pi_{np}^3+\Pi_{nn}^3\big)\nonumber \,, \\  \end{eqnarray}
where the symmetry factors $1/6,1/2,1/2,1/6$  and relative signs are 
already included. The expansion of the (diagonal) polarization functions 
$\Pi_{nn,pp}(\omega,\vec q\,)$ in the isospin-asymmetry parameter $\delta$ takes 
the form (setting $|\vec q\,| = 2 s k_f$ and $\omega = 2s \kappa\, k_f^2/M$):
\begin{equation}\Pi_{nn,pp}(\omega,\vec q\,) = {M k_f \over 4\pi^2 s} \Big\{ 
Q_0(s,\kappa) \pm {\delta \over 6} Q_0'(s,\kappa)+ {\delta^2 \over 36}  
Q_0''(s,\kappa)+\dots \Big\}\,, \end{equation}  
with the linear and quadratic correction functions:
\begin{equation} Q_0'(s,\kappa)=\ln{(1+s)^2+\kappa^2 \over (1-s)^2+\kappa^2 }\,,
\end{equation} 
\begin{equation}Q_0''(s,\kappa)={4s(s^2+\kappa^2-1) \over [(1+s)^2+\kappa^2] 
[(1-s)^2+\kappa^2]}-\ln{(1+s)^2+\kappa^2\over(1-s)^2+\kappa^2 }\,.\end{equation} 
The $pn$-mixed polarization function introduced by the diagram $IV= - exc^3$ has 
the following representation supplemented by an expansion in powers of $\delta/6$: 
\begin{eqnarray}\Pi_{pn}(\omega,\vec q\,) =\Pi_{np}^*(\omega,\vec q\,) &=& 
\int{d^3 l \over (2\pi)^3} {\theta(k_n-|\vec l-\vec q/2|)-\theta(k_p-|\vec l+
\vec q/2|)\over i \omega +(k_n^2-k_p^2+2\vec l \!\cdot\!\vec q\,)/2M}\nonumber 
\\ &=& {M k_f \over 4\pi^2 s} \Big\{ 
Q_0(s,\kappa)- i\, {\delta \over 6} {\widetilde Q}_0(s,\kappa)+ {\delta^2 \over 
36} {\widetilde{\widetilde Q}}_0(s,\kappa)+\dots\Big\}\,. \end{eqnarray}  
One recognizes the difference of neutron and proton chemical potentials 
$(k_n^2-k_p^2)/2M$ in the energy denominator. The pertinent linear and 
quadratic correction functions read:
\begin{equation} \widetilde Q_0(s,\kappa)={\kappa \over s}\ln{(1+s)^2+\kappa^2 
\over (1-s)^2+\kappa^2 }\,,\end{equation}
\begin{equation}{\widetilde{\widetilde Q}}_0(s,\kappa)={4[(1-s^2)^2+(3+s^2)
\kappa^2] \over s[(1+s)^2+\kappa^2] [(1-s)^2+\kappa^2]}-\Big(1+{1\over s^2}\Big) 
\ln{(1+s)^2+\kappa^2\over(1-s)^2+\kappa^2 }\,,\end{equation}
Inserting all these ingredients into eq.(97) and expanding up to order $\delta^2$, 
one obtains for the energy per particle:  
\begin{equation} \bar E(k_f) = 1.0481446 \,{(a_s+a_t)k_f^5 \over \pi^4 M}(5a_s^2
+5a_t^2-14 a_s a_t)\,,\end{equation}
with the numerical coefficient ${\cal N}_1 /4 =1.0481446$ (see eq.(15)).
At the same time the three-ring contribution to the isospin-asymmetry energy 
$A(k_f)\sim k_f^5$ comes out as:
\begin{equation} A(k_f) = {k_f^5 \over \pi^4 M}\big(3.5124\,a_s^3+11.092\,a_s^2 
a_t +10.137\, a_s a_t^2 -5.1014\, a_t^3\big)\,.\end{equation}
The numerical coefficients stem from evaluating the following (convergent) 
double-integrals:
\begin{equation} {1\over 12} \int\limits_0^\infty ds\int\limits_0^\infty d\kappa 
\, Q_0\Big(14 Q_0 Q_0'' +Q_0 {\widetilde{\widetilde Q}}_0+6 Q_0'^{\,2}-
{\widetilde Q}_0^2\Big) = 3.5124\,,\end{equation}
\begin{equation} {1\over 4} \int\limits_0^\infty ds\int\limits_0^\infty d\kappa 
\,Q_0\Big(3Q_0 {\widetilde{\widetilde Q}}_0 -12Q_0 Q''_0+4 Q_0'^{\,2}-3
{\widetilde Q}_0^2\Big)= 11.092\,,\end{equation}
\begin{equation} {1\over 4} \int\limits_0^\infty ds\int\limits_0^\infty d\kappa 
\,Q_0\Big(2Q_0'^{\,2}+3{\widetilde Q}_0^2-6Q_0 Q_0'' -3Q_0 {\widetilde{\widetilde 
Q}}_0\Big)= 10.137\,,\end{equation}
\begin{equation} {5\over 4} \int\limits_0^\infty ds\int\limits_0^\infty d\kappa 
\, Q_0\Big(Q_0 {\widetilde{\widetilde Q}}_0-{\widetilde Q}_0^2\Big)= -5.1014\,.
\end{equation}
When continuing the expansions in eqs.(97,98,101) up to order $\delta^4$, one 
encounters for the quartic isospin-asymmetry energy a coefficient that is 
represented by a logarithmically divergent integral. This feature indicates that 
third-order particle-hole ring diagrams generate the non-analytical term 
$\delta^4 \ln|\delta|$ in the isospin-asymmetry expansion, as it has first been 
found in a second-order calculation with $V_{\rm ct}$ in ref.\cite{quartic}. 
 
\section*{Acknowledgements}
I thank S. Petschauer for support in some calculations and J.W. Holt for 
informative discussions.

\end{document}